# Informed Source Separation using Iterative Reconstruction

Nicolas Sturmel, *Member, IEEE,* Laurent Daudet, *Senior Member, IEEE,*

*Abstract*—This paper presents a technique for Informed Source Separation (ISS) of a single channel mixture, based on the Multiple Input Spectrogram Inversion method. The reconstruction of the source signals is iterative, alternating between a time-frequency consistency enforcement and a re-mixing constraint. A dual resolution technique is also proposed, for sharper transients reconstruction. The two algorithms are compared to a state-of-the-art Wiener-based ISS technique, on a database of fourteen monophonic mixtures, with standard source separation objective measures. Experimental results show that the proposed algorithms outperform both this reference technique and the oracle Wiener filter by up to 3dB in distortion, at the cost of a significantly heavier computation.

## I. Introduction

Audio source separation has attracted a lot of interest in the last decade, partly due to significant theoretical and algorithmic progress, but also in view of the wide range of applications for multimedia. Should it be in video games, web conferencing or active music listening, to name but a few, extraction of the individual sources that compose a mixture is of paramount importance. While blind source separation techniques (e.g. [1]) have made tremendous progress, in the general case they still cannot guarantee a sufficient separation quality for the above-noted applications when the number of sources gets much larger than the number of audio channels (in many cases, only 1 or 2 channels are available). The recent paradigm of Informed Source Separation (ISS) addresses this limitation, by providing to the separation algorithm a small amount of extra information about the original sources and the mixing function. This information is chosen at the encoder in order to maximize the quality of separation at the decoder. ISS can then be seen as a combination of source separation and audio coding techniques, taking advantage of both simultaneously. Actually, the challenge of ISS is to find the best balance between the final quality of the separated tracks and the amount of extra information, so that is can easily be transmitted alongside the mix, or even watermarked into it.

Techniques such as [2], [3], [4] for stereo mixtures, and [5], [6], also applicable to monophonic mixtures, are all based on the same principle: coding energy information about each source in order to facilitate the posterior separation. Sources are then recovered by adaptive filtering of the mixture. For the sake of clarity, we will assume a monophonic case, in a linear and instantaneous mixing (further extensions will be discussed in the discussion Section) : $J$ sources $s_j(t)$, $j = 1 \ldots J$, are linearly mixed into the mix signal $m(t) = \sum_j s_j(t)$. If the local time-frequency energy of all sources is known, noted $|S_k(f,t)|^2$, $k = 1 \ldots J$, then the individual source $s_j(t)$ can be estimated from the mix $m(t)$ using a generalized time-frequency Wiener filter in the Short-Time Fourier Transform (STFT) domain. Computing the Wiener filter $\alpha_j$ of source $j$ is equivalent to computing the relative energy contribution of the source with respect to the total energy of the sources. At a given time-frequency bin $(t, f)$, one has :

$$\alpha_j(t,f) = \frac{|S_j(t,f)|^2}{\sum_k |S_k(t,f)|^2} \quad (1)$$

The estimated source $\tilde{s}_j(t)$ is then computed as the inverse STFT (*e.g.*, with overlap-add techniques) of the weighted signal $\alpha_j(t,f)M(t,f)$, with $M$ the STFT of the mix $m$.

This framework has the advantage that, by construction, the filters $\alpha_j$ sum to unity, and this guarantees that the so-called re-mixing constraint is satisfied :

$$\sum_j \tilde{s}_j(t) = m(t). \quad (2)$$

The main limitation, however, is in the estimation of the phase: only the magnitude $\tilde{S}_j(t,f)$ of each source is estimated by this adaptive Wiener filter, and the reconstruction uses the phase of the mixture. While this might be a valid approximation for very sparse sources, when 2 sources, or more, are active in the same time-frequency bin, this leads to biased estimations, and therefore potentially audible artifacts.

In order to overcome this issue, alternative source separation techniques have been designed [7], [8], taking advantage of the redundancy of the STFT representation. They are based on the classical algorithm of Griffin and Lim (G&L)[9], that iteratively reconstructs the signal knowing only its magnitude STFT. Again, these techniques only use the energy information of each source as prior information, but perform iterative phase reconstruction. For instance, the techniques developed in [7], [8] are shown to outperform the standard Wiener filter. However, in return, reconstructing the phases breaks the re-mixing constraint (2).

The goal of this paper is to propose a new ISS framework, based on a joint estimation of the source signals by an iterative reconstruction of their phase. It is based on a technique called Multiple Input Spectrogram Inversion (MISI) [10], that at each iteration distributes the remixing error $e = m(t) - \sum_j \tilde{s}_j$ amongst the estimated sources and therefore enforces the remixing constraint. It should be noted that, within the context of ISS, it uses the same prior information (spectrograms[1], or quantized versions thereof) as the classical Wiener estimate.

---
[1]The word spectrogram is used here to refer to the squared magnitude of the STFT

Therefore, the results of the oracle Wiener estimate will be used as baseline throughout this paper, "oracle" meaning here with perfect (non-quantized) knowledge of the spectrogram of every source.

In short, the two main contributions of this article can be summarized as follows :

- the modification of the MISI technique to fit within a framework of ISS. The original MISI technique [10] benefits from a high overlap between analysis frames (typically 87.5 %), and the spectrograms are assumed to be perfectly known. The associated high coding costs are not compatible with a realistic ISS application, where the amount of side information must be as small as possible. We show that a controlled quantization, combined with a relaxed distribution of the remixing error, leads to good results even at small rates of side information.
- a dual-resolution technique that adds small analysis windows at transients, significantly improving the audio quality where it is most needed, at the cost of a small - but controlled - increase of the amount of side information.

All these experimental configurations are evaluated for a variety of musical pieces, in a context of ISS.

The paper is organized as follows: a state of the art is given in Section II, where the G&L and MISI techniques are presented. In Section III, we propose an improvement to MISI, with preliminary experiments and discussion. In Section IV, we address the problem of transients and update our method with a dual-resolution analysis. In Section V, the full ISS framework is presented, describing both coding, decoding and reconstruction strategies. Experimental results are presented in Section VI, with a discussion on various design parameters. Finally, Section VII concludes this study.

## II. STATE OF THE ART

### A. Signal reconstruction from magnitude spectrogram

By nature, an STFT computed with an overlap between adjacent windows is a redundant representation. As a consequence, any set of complex numbers $S \in \mathbb{C}^{M \times N}$ does not systematically represent a real signal in the time-frequency (TF) plane. As formalized in [11], the function $\mathcal{G} = STFT[STFT^{-1}[.]]$ is not a bijection, rather a projection of a complex set $S \in \mathbb{C}^{M \times N}$ into the sub-space of the so-called "consistent" STFTs, which are the TF representations that are invariant trough $\mathcal{G}$.

The G&L algorithm [9] is a simple iterative scheme to estimate the phase of the STFT from a magnitude spectrogram $|S|$. At each iteration $k$, the phase of the STFT is updated with the phase of the consistent STFT obtained from the previous iteration, leading to an estimate:

$$\hat{S}^{(k)} = \mathcal{G}(|S|e^{i\angle \hat{S}^{(k-1)}})$$

It is shown in [9] that each iteration decreases the objective function

$$d(\hat{S}^{(k)}, S) = \frac{\sum_{m,n} ||\hat{S}^{(k)}(m,n)| - |S(m,n)||^2}{\sum_{m,n} |S(m,n)|^2} \quad (3)$$

However, this algorithm has intrinsic limitations. Firstly, it processes the full signal at each iteration, which prevents an online implementation. This has been addressed in other implementations based on the same paradigm, see e.g. Zhu et al. [12] for online processing and LeRoux et al. [11] for a computational speedup. Secondly, the convergence of the objective function does not guarantee the reconstruction of the original signal, because of phase indetermination. The reader is redirected to [13] for a complete review on iterative reconstruction algorithms and their convergence issues.

### B. Re-mixing constraint and MISI

In an effort to improve the convergence of the reconstruction within a source separation context, Gunawan et al. [10] proposed the MISI technique, that extracts additional phase information from the mixture. Here, the estimated sources should not only be consistent in terms of time-frequency (TF) representation, they should also satisfy the re-mixing constraint, so that the re-mixing of the estimated sources is close enough to the original mixture. Let us consider the time-frequency remixing error $E_m$ so that:

$$E_m = M - \sum_i \hat{S}_i \quad (4)$$

Note that $E_m = 0$ when using the Wiener filter. In the case of an iterative G&L phase reconstruction, $E_m \neq 0$ at any iteration. Here, MISI distributes the error equally amongst the sources, leading to the corrected source at iteration k, $C_j^{(k)}$:

$$C_j^{(k)} = \mathcal{G}(\hat{S}_j^{(k-1)}) + \frac{E_m}{J} \quad (5)$$

where $J$ is the number of sources.

Therefore, if the spectrogram of the source is perfectly known, it only consists in adapting the G&L technique with an additional phase update based on the re-mixing error:

$$\hat{S}_j^{(k)} = |\hat{S}_j^{(0)}|e^{i\angle C_j^{(k)}} \quad (6)$$

and the MISI algorithm alternates steps 4, 5 and 6. It should be emphasized that, with MISI, the time-domain estimated sources do *not* satisfy the remixing constraint (equation (2)), step (4) playing a role only in the estimation of the phase.

## III. ENHANCING THE ITERATIVE RECONSTRUCTION

The MISI technique [10] presented in the previous section assumes that the spectrogram of every source is perfectly known. However, in the framework of ISS, we have to transmit the spectrogram information of each source with a data rate that is as small as possible, *i.e.* with quantization. At low bit rates (coarse quantization), the spectrograms may be degraded up to the point that modulus reconstruction is necessary. Therefore we will not only perform a phase reconstruction as in MISI, but a full TF reconstruction (phase and modulus) from the knowledge of both the mixture and the degraded spectrogram.

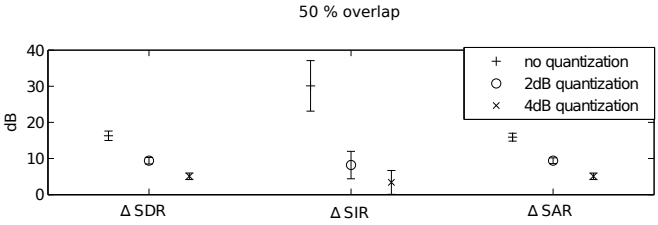

Fig. 2. MISI separation results on the test signal, for different spectrogram quantization levels. Scores are relative to the oracle Wiener filter, and error bars indicate standard deviations.

## A. Activity-based error distribution

It is here assumed that only a degraded version of the source spectrogram is given. Equation (5) can still be used to rebuild both magnitude and phase of the STFT. However, a direct application of this technique leads to severe crosstalk, as some re-mixing error gets distributed on sources that are silent.

In order to only distribute the error where needed, we define a TF domain where a source is considered active based on its normalized contribution $\alpha_j$, as given by the Wiener estimate in eqn. 1. For the source $j$, the activity domain $\Psi_j$ (equation (7)) is the binary TF indicator where the normalized contribution $\alpha_j$ of a source $j$ is above some activity threshold $\rho$

$$\Psi_j(n,m) = \begin{cases} 1 & \text{if } \alpha_j(n,m) > \rho \\ 0 & \text{otherwise} \end{cases} \quad (7)$$

Now, the error is distributed only where sources are active:

$$\hat{S}_j^{(k)}(n,m) = \Psi_j(n,m) \left( \mathcal{G}(\hat{S}_j^{(k-1)}) + \frac{E_m(n,m)}{D(n,m)} \right) \quad (8)$$

where $D(n,m)$ is a TF error distribution parameter. It is possible to compute $D(n,m)$ as the number $N_a$ of active sources at TF bin $(n,m)$ (i.e., $D(n,m) = \sum_j \Psi_j(n,m)$). However, it was noticed experimentally that a fixed $D$ such that $D >> N_a$ provides better results. This means that only a small portion of the error is added at each iteration, and that the successive TF consistency constraint enforcements (the $\mathcal{G}$ function) validate or invalidate the added information. The exact tuning of parameters $D$ and $\rho$ is based on experiments, as discussed in section III-B. We expect that the lower $\rho$, the lesser the artifacts of the reconstruction, but also the higher the crosstalk (sources interferences) because the remixing error is distributed on a higher number of bins.

## B. Preliminary experiments

A first test is performed to validate the proposed design, and to experiment on the various parameters. We use a monophonic music mixture of electro-jazz at a 16bits/44.1kHz format. Five instruments are playing in this mixture : a bass, a drum set, a percussion, an electric piano and a saxophone. These instruments present characteristics that interfere with one another. For instance, the bass guitar and the electric piano are heavily interfering in low frequencies, whereas drums and percussions both have strong transients. The saxophone is very breathy but the breath contribution is far below the energy of the harmonics.

The spectrograms are log-quantized (in dB, cf [14], [6]) with three quantization steps : $u = 0$ (no quantization), 2 and 4dB. For each of these three conditions, we use two overlap values of 50% and 75% and a window size of 2048 samples at 44,1kHz sampling rate. Two values of the activity threshold are tested: $\rho = .1$ and $.01$. The phase of each source is initialized with the phase of the mixture, and 50 iterations were performed.

We test 3 variants of the proposed separation method :
1) M1 : with $D = 40$ and activity detection.
2) M2 : with $D = N_a$ and activity detection.
3) M3 : with $D = N_a$ and no activity detection.

For this evaluation, we use the three objective criteria of the BSS Eval toolbox[15], namely the Source to Distortion Ratio (SDR), the Source to Interference Ratio (SIR) and the Source to Artifact Ratio (SAR). Results given on Figure 1 are relative to the Oracle Wiener filter estimation performances, taken as reference. In the present experiment the absolute mean (respectively, standard deviation) of the Oracle Wiener filter were : SDR = 9.0 (1.3) $dB$, SIR = 21 (5.1) $dB$, SAR = 9.4 (1.2) $dB$ for both 50% and 75% overlap. Results of MISI on the same signal are given on Figure 2.

## C. Discussion

The results are presented on Figures 1 and 2 and the reconstructed sources are available on the demo webpage [16]. The performance of unquantized MISI is very high, but decreases rapidly when quantization increases. This is directly linked to the fact that the spectrogram is constrained, which would be even more problematic when part of this spectrogram is missing, for bitrate reduction purposes. The activity-based error distribution (M1 and M2 vs M3) improves significantly the three objective criteria both in mean and standard deviation. This is expected as the activity domain prevents reconstruction of a source on a bin where its contribution to the mixture is negligible. One can also see that lowering the activity threshold $\rho$ (from .1 - upper line - to .01 - lower line -) improves the SAR but lowers the SIR: a lower value of $\rho$ distributes the error on a larger amount of bins. While this provides less "holes" in the reconstructed TF representation (higher SAR), it also involves more crosstalk between sources (lower SIR). In every condition, the tradeoff between SIR and SAR when lowering $\rho$ seems to be a loss of about 1dB on the SIR for a gain of 1dB on the SAR. Since the SIR is already high on the oracle Wiener filter ($> 15$dB), it seems a better tradeoff to favor SAR, in order to improve the global SDR gain. Therefore, the lower value $\rho = .01$ will be used for the rest of the paper.

The improvements brought by $D >> N_a$ (M1) compared to $D = N_a$ (M2) are less important. The precise choice of $D$ is experimented on fig. 3. Large values of $D$ seem to provide a better convergence: the energy of the error that is distributed to a source but that does not belong to it (on a consistency basis) will be easily discarded because of its small value and because of the energy smearing effect of the $\mathcal{G}$ function.

When the spectrogram is quantized with $u = 4$ dB quantization step, the reconstruction performance reaches a maximum with $D = 40$ for 50 iterations.

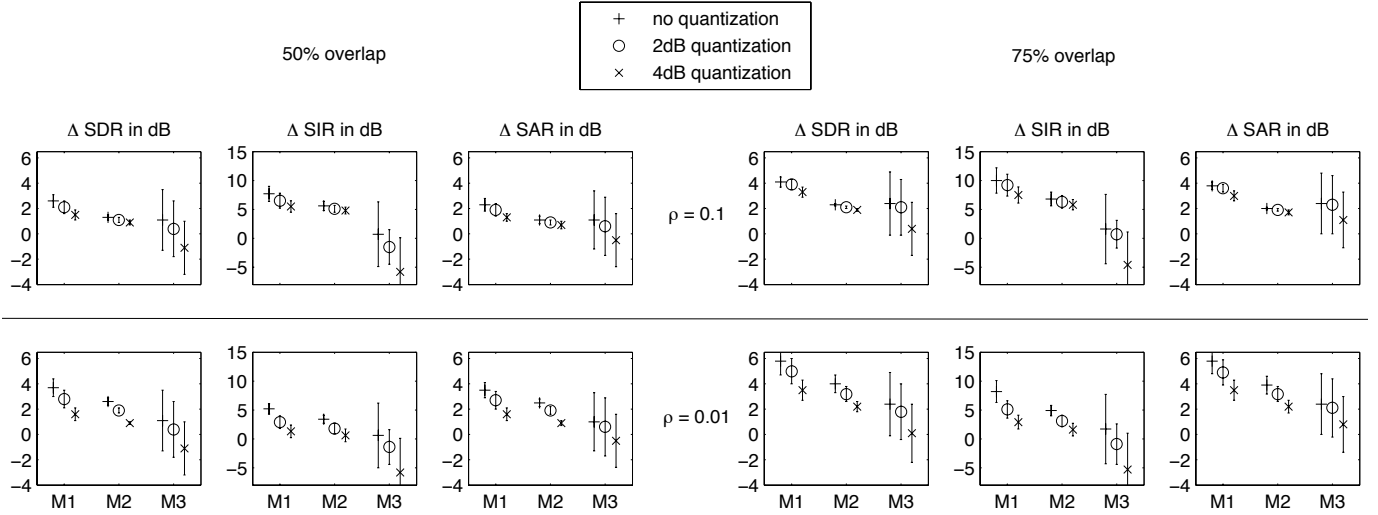

Fig. 1. Separation results for the three variants of the proposed method : M1 ($D = 40$, activity detection), M2 ($D = N_a$, activity detection) and M3 ($D = N_a$, no activity detection). Scores are relative to the oracle Wiener filter, and error bars indicate standard deviations. Different parameters are tested : the quantization step $u$, the STFT overlap (50% and 75%) and the activity threshold $\rho$.

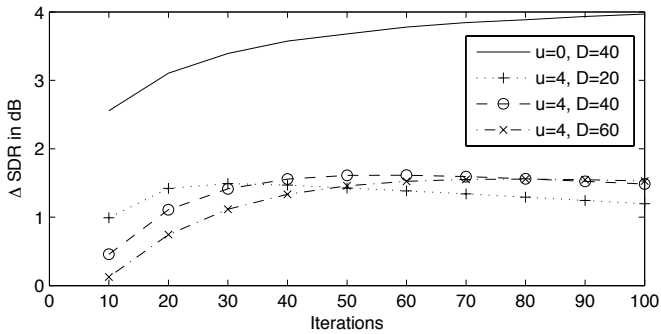

Fig. 3. Different values of $D$ for different number of iterations.

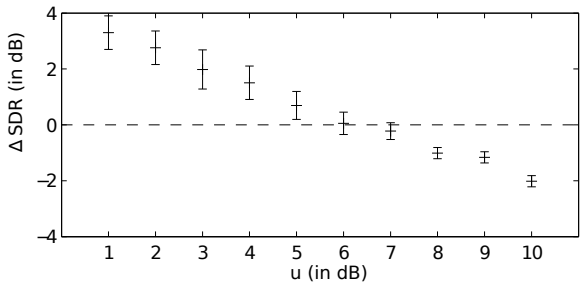

Fig. 4. Improvements over the Wiener filter for a varying quantization step $u$. Window size of 2048 samples with 50% overlap, $\rho = 0.01$.

Finally, the effect of spectrogram quantization is clear. As expected, increasing the quantization steps lowers the SDR but also dramatically lowers the SAR because of added artifacts caused by the quantization. Figure 4 presents the SDR improvement when varying the quantization step $u$, for algorithm M1. Even for a relatively high quantization step of 4dB, results still outperform the oracle Wiener filter.

To summarize the results of this preliminary experiment, we have shown that - at least for the sounds under test - the proposed method M1 (activity detection, $D = 40$) can outperform the oracle Wiener filter, while keeping the amount of side information low, with a crude quantization of the spectrograms ($u = 4$ dB). However, these results are not perfect, especially in terms of perception. When listening to the sound examples (available online [16]), one can hear a number of artifacts, especially at transients. Indeed, transient reconstruction from a spectrogram or from a Wiener filter is a well-known issue [17], as time domain localization is mainly transmitted by the phase. The next section alleviates this problem by using multiple analysis windows.

## IV. IMPROVING TRANSIENTS RECONSTRUCTION

The missing phase information at transients leads to a smearing of the energy, pre-echo or an impression of over smoothness of the attack. In order to prevent these issues, a window switching can be used, with shorter STFT at transients [17], [18], [19]. In Advanced Audio Coding (AAC) for instance, the window switches from 2048 to 256 samples when a transient is detected. Here, because we want the same TF grid for sources that can have very different TF resolution requirements, we do not switch between window sizes but rather use a dual resolution at transients, keeping both window sizes. Note that this leads to a small overhead in terms of amount of side information to encode (both short- and long-window spectrograms have to be quantized and transmitted at transients), but does not require transition windows.

### A. Transients detection

We use the same non-uniform STFT grid for every source and for the mixture, keeping the ability of TF addition and subtraction for error distribution. In order to obtain this non-uniform grid, we process in three steps at the coding stage:
1) a binary transient indicator $T_j(t)$ is computed for each source $j$, using the Complex Spectrum Difference [20]:

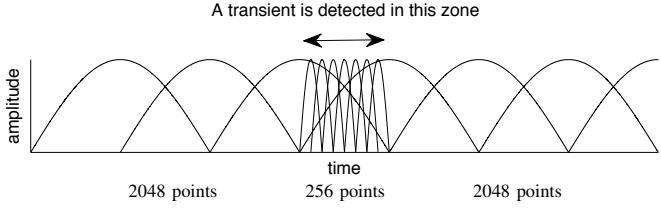

Fig. 5. Large and small windows in the dual-resolution STFT.

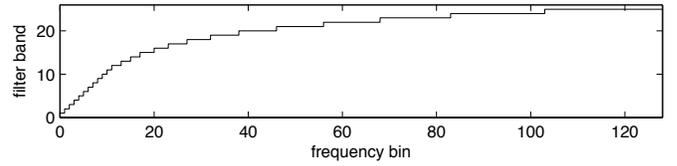

Fig. 7. Logarithmic bin grouping in subbands, for 25 subbands and 129 bins

$T_j$ equals to 1 if a transient is detected at time $t$, 0 otherwise.

2) The transients are combined in $T_{all}$ so that

$$T_{all} = T_1 \oplus T_2 \oplus T_3 \oplus \ldots$$

where $\oplus$ is the logical OR function.

3) $T_{all}$ is cleaned so that the time between two consecutive transients is greater or equal to the length of two large windows.

The non-uniform STFT is therefore constructed by concatenation of the large-window STFT on all frames, plus of short-window STFT on transient frames in $T_{all}$. Figure 5 shows this dual-resolution STFT when a transient is detected.

### B. Experiments

In order to evaluate the improvements brought by dual-resolution, we use the same sound samples as before : an electro-jazz piece of 15 seconds of music composed of 5 sources. The same parameters are also used: 50 iterations, $D = 40$, $\rho = 0.01$, and two overlap values : 50% and 75%. The large and small window sizes are set to 2048 and 256 samples, respectively.

Results are presented on Figure 6, showing improvement over the Wiener filter as before. Note that we used the same Wiener filter reference (single-resolution) throughout this experiment. Transient detection with 50% overlap (leading to an increase in data size from 15 to 25%, depending on the number of detected transients), are close to the results obtained with an uniform STFT at 75% overlap (100% more data): transient detection brings the same separation benefits as increasing the overlap, with the added value of sharper transients. Audio examples are available on the demo web page [16].

## V. PRACTICAL IMPLEMENTATION IN AN ISS FRAMEWORK

This section presents the new source reconstruction method in a full ISS framework. We call our method Informed Source Separation using Iterative Reconstruction (ISSIR). First the coding scheme will be presented, together with parameter tuning. Then, the decoding scheme will be presented.

### A. Coder

Data coding is used to format and compact the information needed for the posterior reconstruction. The size of this coded data is of prime importance :

- In the case of watermarking within the mixture (which would then be coded in PCM), high capacity watermarking may be available [21], limited by a constraint of perceptual near-transparency. The lower the bit rate, the higher the quality of the final watermarked mixture, used for the source reconstruction.
- In the case of a compressed file format for the mixture, the side-information could be embedded as meta-data (AAC allows meta-data chunks, for instance). In this case, the size of the data is also important in order to keep the difference between the coded audio file and the original audio file to a minimum.

Of course, increasing the bit rate would eventually lead to the particular case where simple perceptual coding of all the sources (for instance with MPEG 2/4 AAC) would be more efficient than informed separation.

In order to achieve optimal data compaction, we make the following observation: most of the music signals are sparse and mostly described by their most energetic bins. Therefore, spectrograms coding should not require the description of TF bins with an energy threshold $T$ lower than e.g. -20dB below the maximum energy bin of the TF representation. What we propose is then to discard the bins that are lower than $T$ in Energy. $T$ is the first parameter to be adjusted in order to fit the target bit rate, with $T \leq -20$ dB. Note that former work, e.g. [6], also threshold the spectrogram, but much lower in energy (-80 dB). The second parameter for data compaction is the quantization of the spectrogram with step $u$. As seen before, increasing $u$ decreases the reconstruction quality but lowers the number of energy levels to be encoded. Since increasing $u$ did not change much the entropy of the data distribution, we choose $u = 1 dB$ for the whole experiment. The third parameter $\rho$ used for the activity domain is set to .01 and is not modified in our experiments.

The data size of the activity domain is then fixed throughout the experiments. In order to compact this information even more, we group time-frequency bins on the frequency scale using logarithmic rules similar to the Equivalent Rectangular Bandwidth (ERB [22]) scale. This psychoacoustic-based compression technique has also been used in informed source separation in [4], [5]. For the experiments in this paper we use 75, 125 or 250 non overlapping bands on large windows (1025 coded bins) and 25 bands on small windows (129 coded bins), as presented on Figure 7.

Additional parameters such as spectrogram normalization coefficients, STFT structure, transient location and quantization step are transmitted apart: such information represents a negligible amount of data as most of it is fixed for the whole file duration. At the end of the coding stage, a basic entropy coding (in our experiment setup, we used bzip2) is added.

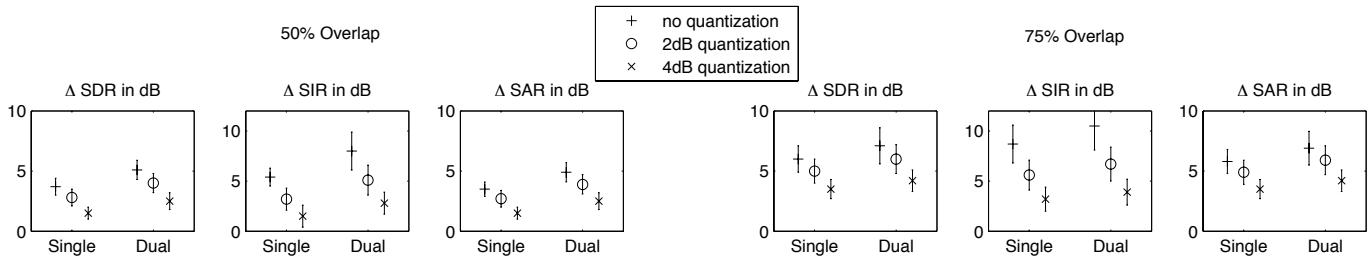

Fig. 6. Separation results for a single resolution STFT (*single*) vs. dual-resolution STFT (*dual*), for various quantization steps $u$ and 2 overlaps.

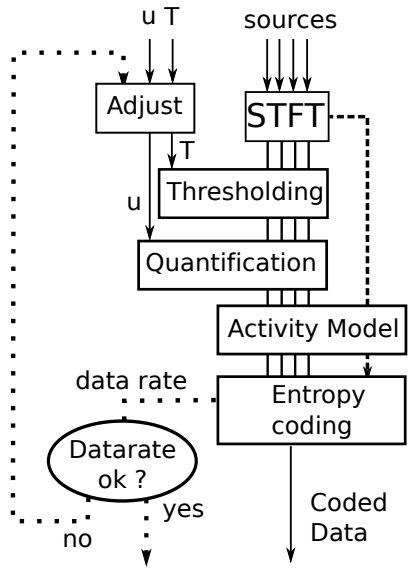

Fig. 8. Block diagram of the ISSIR coding framework.

Figure 8 shows the coding scheme, with the feedback loop for the adjustment of the model parameters to the target bit rate in kb/source/s. The target bit rate is a mean amongst the sources, as some sources will require more information to be encoded than others. Such framework allows mean data rates as low as 2kb/source/s.

### B. Decoder

The decoder performs all the previous operations backwards. It first initializes each source using the log-quantized data and the phase of the mixture $M$. Then, the iterative reconstruction is run for K iterations and the signals are finally reconstructed using the decoded activity domain $\Psi_j$.

## VI. EXPERIMENTS

In this section we validate our complete ISSIR framework on different types of monophonic mixtures. As the problem of informed source separation is essentially a tradeoff between bit rate and quality, we perform the experiments by setting different thresholds $T$ and filter bank sizes for the single and dual window STFT algorithm presented before. The baseline for comparison is a state-of-the-art ISS framework based on Wiener filtering [6], where JPEG image coding is simply used to encode the spectrograms. For a fair comparison, we also use this method with the same ERB-based filter bank grouping. For reference, we also compute the results of the original MISI method, with spectrogram quantization and coding.

The test database is composed of 14 short monophonic mixtures from the Quaero database[2], from 15 to 40 s long, with various musical styles (pop, rock, industrial rock, electro jazz, disco) and different instruments. Each mixture is composed of 5 to 10 different sources, for a total of 90 source signals. The relation between the sources and the mixture is linear, instantaneous and stationary ; however, the sources include various effects such as dynamic processing, reverberation or equalization, so that the resulting mixtures are close to what would have been obtained by a sound engineer on a Digital Audio Workstation.

Figure 9 presents the mean and standard deviation of the improvements over the oracle Wiener filter for the whole database. As before, SDR, SIR and SAR are used for the comparison of the different methods. Reported bit rates are averaged over the whole database, at a given experimental condition. Four mixtures under Creative Commons license are given as audio examples on the demo web page [16]:

- Arbaa (Electro Jazz) - mixture nb. 2 - 5 sources
- Farkaa (Reggae) - mixture nb. 4 - 7 sources
- Nine Inch Nails (Industrial Rock) - mixture nb. 8 - 7 sources
- Shannon Hurley (Pop) - mixture nb. 12 - 8 sources

### A. Bit rates and overall quality

As expected, increasing the bit rate improves the reconstruction on all criteria. The two ISSIR algorithms always outperform the baseline method of [6], although not significantly at very low bit rates when the non-uniform filterbank is used. The dual-resolution framework requires more data, and only outperforms the single resolution algorithm for bit rates higher than 10kb/source/s, where the latter tends to reach its maximum of 1.7dB improvement over the oracle Wiener filter. At 32kb/source/s, the dual resolution method reaches its own maximum of approx. 3dB improvement over the oracle Wiener filter. For even higher bit rates, MISI gives significantly better results, but the high amount of total side information is not compatible with a realistic ISS usage.

[2]www.quaero.org

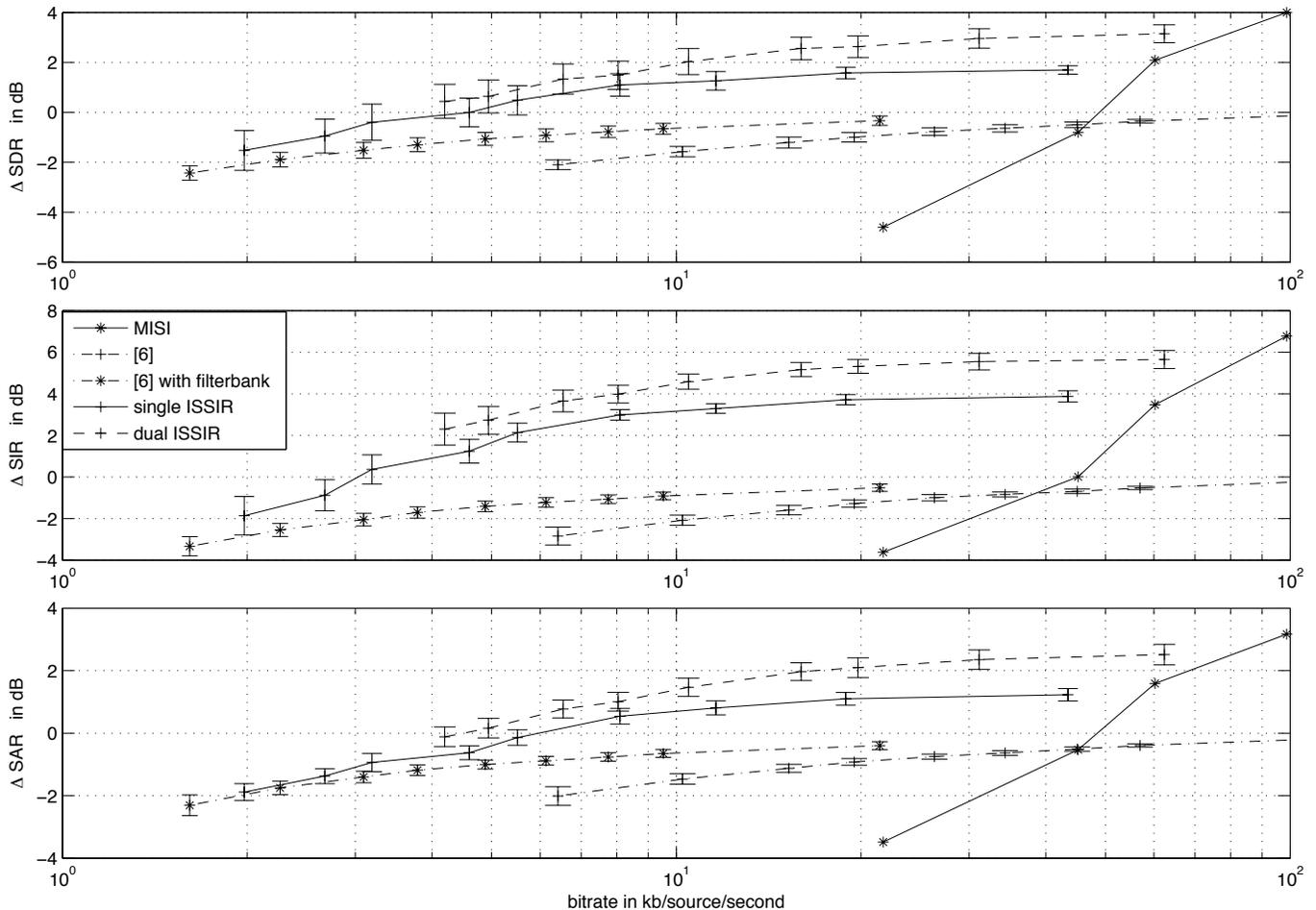

Fig. 9. Reconstruction results for the different methods, on monophonic mixtures at different bit rates. Results are given relative to the oracle Wiener filter.

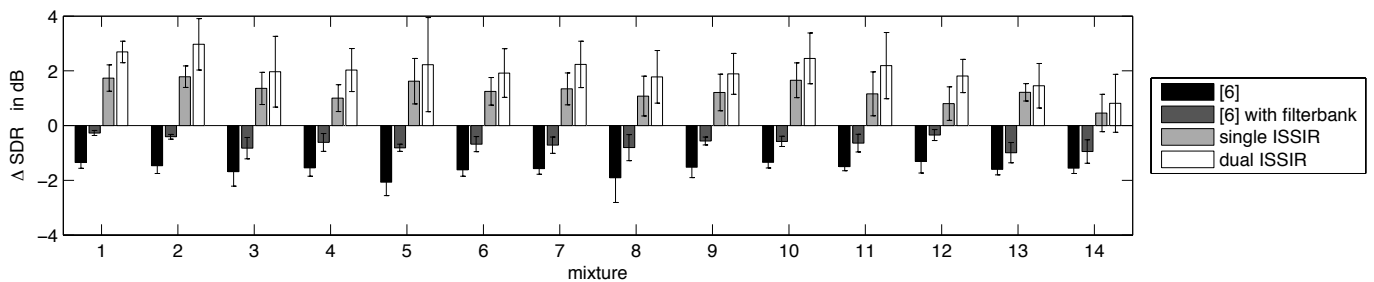

Fig. 10. Separation results compared to the oracle Wiener filter for every tested mixture with mean and standard deviation for a bit rate of 10kb/source/second.

### B. Performance as a function of the sound file

The previous experiments are associated with a strong variance: results are highly dependent both on the type of music and on the sources. Figure 10 presents the SDR results for the 14 sound files, at an average bit rate of 10kb/source/s. It can be observed that the variations are happening both from mixture to mixture and within the mixture. At this bit rate, the dual resolution algorithm may not always perform better than the single resolution algorithms, as can be seen for mixtures 3, 5, 13, and 14. However, the proposed technique (single or dual) always outperforms the reference method of [6].

### C. Computation time

Since the proposed reconstruction algorithm is iterative, the decoding requires a heavier computation load than simple Wiener estimates. A Matlab implementation of the dual-resolution scheme led to computation times of 6 to 9 s per second of signal, for 50 iterations, on a standard computer.

As a proof of concept, the single resolution iterative reconstruction was also implemented in parallel with the OpenCl [23] API, using a fast iterative signal reconstruction [11]. On a medium range graphic card, the computation time dropped to .3 to .4 s per second of signal. The adaptation of this fast scheme to the dual resolution case is, however, not straightforward.

*D. Complex mixtures*

In the case of complex mixtures (multichannel, convolutive, etc), the main issue is the error distribution as in equation (8), that requires itself a partial inversion of the mixing function. In fact, actual source separation is done at this level, and this paper shows that a simple binary mask at this stage is sufficient in order to achieve good results on monophonic mixtures. The framework presented in this paper could then be adapted for a vast variety of source separation methods, especially in the cases when the mixing function is known. In the case of multichannel mixtures, for instance, error repartition distribution be done using beamforming techniques.

## VII. Conclusion

This paper proposes a complete framework for informed source separation using an iterative reconstruction, called Informed Source Separation using Iterative Reconstruction (ISSIR). In experiments on various types of music, ISSIR outperforms on standard objective criteria a state-of-the-art ISS technique based on JPEG compression of the spectrogram, and even the oracle Wiener filtering by up to 3dB in source-to-distortion ratio.

Future work should focus on the optimization of the algorithm in order to lighten the computation load, and on its extension to multichannel and convolutive mixtures. Psychoacoustic models should also be considered as a way to compact and shape the side information. Finally, formal listening tests should confirm the objective results, although it should be emphasized that setting up a whole methodology for such ISS listening tests (that is not established as in other fields, *e.g.*, audio coding), is a work in itself that goes beyond the current study.

## Acknowledgment

This work was supported by the DReaM project (ANR-09-CORD-006) of the French National Research Agency CONTINT program. LD acknowledges a joint position with the Institut Universitaire de France. The authors would like to thank the consortium of the DReaM project for fruitful discussions, and in particular A. Liutkus, G. Richard and L. Girin who provided the test material.